\documentclass[thmsa,a4paper,final]{article}
\usepackage{sw20asas}

\input tcilatex
\QQQ{Language}{
American English
}

\begin{document}

\begin{center}
\textbf{II - LOCAL SOLUTION}

\textbf{of}

\textbf{\ A SPHERICAL HOMOGENEOUS AND ISOTROPIC UNIVERSE}

\textbf{radially }

\textbf{DECELERATED TOWARDS THE EXPANSION CENTER :}

-

\textbf{\ TESTS ON HISTORIC DATA SETS}

\vspace{.7in}
\end{center}

- 1st version: January, 1997 - 2nd ver.: January, 1998 - 3rd ver.: February,
1999 -

by \textbf{Luciano Lorenzi (E-mail: l.lorenzi@senato.it),}

\textbf{former astronomer at the Astronomical Observatory of Turin - Italy -}

\vspace{.7in}

\begin{center}
\textbf{ABSTRACT}
\end{center}

\begin{quote}
\textbf{The topic of the paper is the mathematical analysis of a radially
decelerated Hubble expansion from the Bahcall \&\ Soneira void center (}$%
\alpha \approx 9^h;\delta \approx +30^0$)\textbf{.}

\textbf{Such analysis, based on the result }$K_0$\textbf{\ }$=\left( \delta
H/\delta r\right) _0=3H_0^2/c$\textbf{\ of Paper I, in the hypothesis of
local homogeneity and isotropy, gives a particular Hubble ratio dipole
structure to the expansion equation, whose solution has been studied at
different precision orders and successfully tested on a few historic data
sets, by de Vaucouleurs (1965), by Sandage \& Tammann (1975), and by
Aaronson et al. (1982-86). The fittings of both the separate AA1 and AA2
samples show a good solution convergence as the analysis order increases,
giving even coinciding solutions when applied to 308 nearby individual
galaxies (308AA1) and to 10 clusters (148AA2), respectively.}

\textbf{As a result one obtains:}

$R_0\approx 210$\textbf{\ }$Mpc$\textbf{\ and }$H_0=88.0\pm 1.6$\textbf{\ }$%
Km$\textbf{\ }$s^{-1}Mpc^{-1}$ \textbf{by Aaronson et al., or}

$R_0\approx 260$\textbf{\ }$Mpc$\textbf{\ and }$H_0=70\pm 3$\textbf{\ }$Km$%
\textbf{\ }$s^{-1}Mpc^{-1}$ \textbf{by Sandage \& Tammann.}

\textbf{In conclusion an optional further graphical check of the model is
presented by plotting a series of Hubble ratio average dipoles.}
\end{quote}

\section{INTRODUCTION}

The present paper, as a natural continuation of the previous one (I), has
the function of carrying out a series of cosmological tests at observational
experimental level. The checks have been based on a few very meaningful
historic data samples, in order to show the general faithfulness of the
modelled radial Hubble expansion from the Bahcall and Soneira void center.
Of course the application to more recent data must follow immediately, but
the implications are so crucial that it is still necessary to confirm the
correctness of the model, actually starting from ancient data like those of
the $1960^{\prime }s$ by de Vaucouleurs , continuing through the historic $%
^{\prime }70^{\prime }s$ years measures by Sandage \& Tammann, to the more
recent results of the $^{\prime }80^{\prime }s$ years by Aaronson et al..
All these data refer to a nearby Universe where ''nearby'' here means a
distance range of about 100 Mpc.

\vspace{.4in}

\subsection{Local homogeneity and isotropy}

In this ''nearby'' entourage we have assumed that the Universe is
homogeneous and isotropic, according to the Cosmological Principle. Indeed,
modern Cosmology has been searching for far away homogeneity for a long
time, on a large scale, as the Galaxy environment seems to show too many
irregularities. However, our different starting point is different only in
appearance, as the assumed local homogeneity and isotropy depends on the
adopted spherical model in which homogeneity and isotropy can be averagely
attributed to all the volume containing the huge void of Bahcall \& Soneira,
that is all the inner Universe and its expansion center. Of course, in this
case the effects on our entourage are determined by our inner Universe while
those on more remote clusters should be tied to their larger or smaller
inner Universe. Now, based on the preliminary results referred to on paper
I, the quadrupole space effect $Q$ ($\Delta H_{ga}\rightarrow 0$) seems
negligible for a depth of about 100 Mpc; and this means the adoption of the
homogeneity and anisotropy is the correct condition to study the samples of
the nearby Universe.

\newpage\ 

\subsection{Apparent anisotropy}

On the grounds of the fundamental expansion equation obtained in paper I:

\begin{equation}
\dot r=Hr+\Delta H\cdot (r-R\cos \gamma )  \label{1}
\end{equation}
and of the basic assumption of homogeneous and isotropic Universe, the
finite difference $\Delta H$ should be zero in order to avoid any space
dependence. Indeed that will be the case in absence (see paper I) both of
the above cited space effect (SE)($\Delta H_{ga}=0$) and of the other light
time effect (TE)($\Delta H_{MW}=0$); in particular the latter zero TE would
mean that the light propagation is instantaneous, or, that the speed of
light has an infinite value. In this hypothetical context, and only in this,
the observed Universe should look exactly like that foreseen by the pure
Hubble law: $\dot r=Hr$, independently of the modelled decelerating radial
expansion from a definite center. In other words, if the quantities $\dot r$
and $r$ refer to the same moment, the canonical Hubble law holds , and $%
\Delta H$ is zero; otherwise, if $\dot r$ is the observed $\dot r_{obs}$ ,
its value will be affected by two different epochs \textbf{due to the light
delay linked to our decelerated Galaxy}, while $r$ as light space distance
will represent only the emission epoch owing to the constancy of the light
speed; and so even a homogeneous and isotropic Universe will have to look
like that described by the varied Hubble law of Eq. (1). Hence, the adopted
status of homogeneity and isotropy must refer to the same instant, or epoch,
of our proper time, while the observed anisotropy according to Eq. (1) with $%
\Delta H=\Delta H_{MW}$, in spite of its strong observational evidence,
practically is only apparent and due to a prospect effect generated by the
light delay.

\newpage\ 

\section{ANALYSIS}

\subsection{THE EXPANSION EQUATION ACCORDING TO $K_0=3H_0^2/c$}

In order to solve the fundamental expansion Eq. (1), once accepted the
experimental result $K_0=3H_0^2/c$ of Paper I in the assumed case of
homogeneous and isotropic spherical universe, one needs to substitute $H,$ $%
\Delta H,$ $R$ with the formulas (38)(40) by Paper I . The result is:

\begin{equation}
\frac{\dot r_{obs}}r=\frac{H_0c}{c-3H_0r}+\frac{3H_0^2}{c-3H_0r}\left[
r-R_0\left( 1-\frac{3H_0r}c\right) ^{\frac 13}\cos \gamma \right]  \label{2}
\end{equation}

that, after a short processing, takes the following useful formulation:

\begin{equation}
\frac{\dot r_{obs}}r=H_0\left( \frac{1+x}{1-x}\right) -K_0R_0\cdot \left(
1-x\right) ^{-\frac 23}\cos \gamma  \label{3}
\end{equation}

where there are: 
\begin{eqnarray*}
\frac{\dot r_{obs}}r &:&\text{ variable quantity as Hubble ratio}  \label{4}
\\
H_0 &:&\text{ Hubble Constant} \\
K_0R_0 &=&\frac{3c}2\frac{2H_0^2R_0}{c^2}=a_0:\text{Galaxy radial
deceleration coefficient (see Eq. (41) in paper I)} \\
x &=&\frac{3H_0r}c:\text{ variable quantity whose value in the nearby
Universe is}\ll 1 \\
\cos \gamma &:&\text{ variable quantity according to the observed position}
\end{eqnarray*}

Furthermore, applying the Mc Laurin Series, we have:

\begin{equation}
\frac{1+x}{1-x}=1+2x+2x^2+.....  \label{4}
\end{equation}
while, according to the binomial series, it is:

\begin{equation}
\left( 1-x\right) ^{-\frac 23}=1+\frac 23x+\frac 59x^2+....  \label{5}
\end{equation}

The previous two series developments allow to study the expansion Eq. (3) at
different precision orders, and consequently to verify the solution
convergence.

\subsubsection{0$th$ order analysis}

The 0$th$ order formulation of Eq. (3) is that following the stopping of the
series at the zero order term, that is at 1. In this extreme case,
corresponding to $r\rightarrow 0$, the expansion equation becomes:

\begin{equation}
\frac{\dot r_{obs}}r=H_0-K_0R_0\cos \gamma  \label{6}
\end{equation}

The aforesaid extremely simplified formula has importance especially in the
graphical applications, when referred only to a very nearby Universe, of
course. Here the Hubble ratios plotted against $\cos \gamma $ have to
dispose themselves according to an \textbf{indisputable dipole structure, }%
displayed by the verified deceleration angular coefficient $K_0R_0=a_0$.

\subsubsection{1$st$ order analysis}

The first order formulation of Eq. (3) is that following the stopping of the
series (4)(5) at the first order term, which are $2x$ and $\frac 23x$
respectively. Here, after the substitution $x=\frac{3H_0r}c$, the resulting
equation is:

\begin{equation}
\frac{\dot r_{obs}}r=H_0\left( 1+\frac{6H_0r}c\right) -a_0\left( 1+\frac{%
2H_0r}c\right) \cos \gamma  \label{7}
\end{equation}
that, in terms of 2nd degree algebraic equation, becomes:

\begin{equation}
\frac{\dot r_{obs}}r+a_0\cos \gamma =H_0\left( 1-2a_0\frac{r\cos \gamma }%
c\right) +\frac{6r}cH_0^2  \label{8}
\end{equation}

Solving the above 2nd degree equation in $H_0$ could mean finding, by the
least square method, the value of the deceleration parameter: 
\begin{equation}
a_0=\frac{3H_0^2R_0}c  \label{9}
\end{equation}
, hence the $R_0$ value corresponding to the resulting $H_0$ , able to
minimize the standard deviation, when an homogeneously distributed sample of 
$\dot r$ and $r$ measures of the nearby Universe is available.

Even the $1st$ order Eq. (7) may have a graphical representation like that
of Eq. (6), but now, being $r=r_{*}$ the average distance of a not dispersed
sample, we shall have a paper I (68)-type formula to represent an\textbf{\
average dipole}, with $H_{*}=H_0+2K_0r_{*}>H_0$ and $R_{*}=R_0(1+\frac{%
2H_0r_{*}}c)>R_0$, that is

\begin{equation}
\frac{\dot r_{obs}}r=H_{*}-K_0R_{*}\cos \gamma  \label{10}
\end{equation}

\subsubsection{N$th$ order analysis}

The N$th$ order formulation is that directly following the expansion
equation (2), now written as:

\begin{equation}
\frac{\dot r_{obs}}r\left( 1-\frac{3H_0r}c\right) =H_0+\frac{3H_0^2}cr-\frac{%
3H_0^2R_0}c\cos \gamma \left( 1-\frac{3H_0r}c\right) ^{\frac 13}  \label{11}
\end{equation}
with the only binomial $\left( 1-\frac{3H_0r}c\right) ^{\frac 13}$developed
by series, according to:

\begin{equation}
\left( 1-\frac{3H_0r}c\right) ^{\frac 13}=1-\frac{H_0}cr-\frac{H_0^2}{c^2}%
r^2-\Sigma  \label{12}
\end{equation}
being 
\begin{equation}
\Sigma =\frac 53\QOVERD( ) {H_0r}{c}^3+\frac{10}3\QOVERD( ) {H_0r}{c}^4+%
\frac{22}3\QOVERD( ) {H_0r}{c}^5+...  \label{13}
\end{equation}

Then Eq. (11), after a short processing with the (9) adoption, gives the
following 2nd degree algebraic equation in $H_0$:

\begin{equation}
H_0^2\left( 3\frac rc+a_0\frac{r^2\cos \gamma }{c^2}\right) +H_0\left( 1+3%
\frac{\dot r}c+a_0\frac{r\cos \gamma }c\right) -\left( \frac{\dot r}%
r+a_0\cos \gamma -a_0\Sigma \cos \gamma \right) =0  \label{14}
\end{equation}

whose solution could mean finding, using the least square method, the value
of the deceleration parameter $a_0$ (9), that is the corresponding $R_0$ ,
able to minimize the standard deviation of the fitting carried out on a
sample of galaxies/groups/clusters/superclusters, having computed $\Sigma $
by successive approximations of $H_0.$

\subsubsection{The q$_0$ test}

In order to check the most correct solution, it is now possible to apply a
powerful test based on the conservation of the parameter $q_0$, that is of 
\begin{equation}
q_0=-\frac{H_0R_0}c  \label{15}
\end{equation}
which, keeping in mind the previous paper I, represents the Galaxy recession
velocity with respect to the expansion center.

Starting from the 0$th$ order equation (6)(but the same result comes from
higher orders), that holds for $r\rightarrow 0$, it is easy to show the $q_0$
conservation, rewriting the (6) as follows: 
\begin{equation}
\dot r_{obs}=H_0r(1+3q_0\cos \gamma )  \label{16}
\end{equation}

As samples may differ practically because of different absolute estimates of
distance, the differential of eq. (16), according to the error propagation,
tells us that the same radial velocity $\dot r_{obs}$ of any group or
galaxy, that is $d\dot r_{obs}=0$ , implies the following logic sequence: 
\begin{equation}
\text{IF }\hspace{.3in}\frac{dR_0}{R_0}=\frac{dr}r\hspace{.3in}\text{THEN}%
\hspace{.3in}\frac{dr}r=-\frac{dH_0}{H_0}=\frac{dR_0}{R_0}\Rightarrow q_0=%
\text{constant}  \label{17}
\end{equation}

\section{RESULTS}

\subsection{PRELIMINARY TEST ON HISTORIC NEARBY SAMPLES}

Having stated this, firstly let's go on to check the 0$th$ and 1$st$ order
solution applying Eqs. (6) and (8) to two historic data samples of very
nearby galaxy Groups, all inside a distance range of 30 Mpc, by means of
least square linear fittings and assuming $\gamma =0^0$ at $\alpha \approx
9^h,$ $\delta \approx +30^0$ or $l\approx 195^0,$ $b\approx +40^0$ or $%
SGL\approx 60^0,$ $SGB\approx -35^0$ (VC coordinates) and $\gamma =180^0$ at 
$\alpha \approx 21^h,$ $\delta \approx -30^0$ (AVC=anti void center). The
group position is considered applying the trigonometrical function cosine of
the observed angle $\gamma $ between the direction of VC and that of the
galaxy group: 
\begin{equation}
\cos \gamma =\sin \delta _{VC}\sin \delta +\cos \delta _{VC}\cos \delta \cos
(\alpha -\alpha _{VC})  \label{18}
\end{equation}
The samples are the following:

\subsubsection{Groups by de VAUCOULEURS}

DV sample:- 52 nearby Groups of galaxies (368 in all) , whose preliminary
elements had been reported by de Vaucouleurs (1965-Table 2) thirty-five
years ago. Weighing each group by the number of galaxies, $w=n_{obs}$ , two
fitting solutions follow, at 0$th$ and 1$st$ order respectively, being $s$
the standard deviation or $s_{MIN}$ the minimum one varying $a_0$; that are: 
\begin{eqnarray}
\text{0}^0 &&\text{: }a_0=23.2\pm 1.8\text{ ; }H_0=108.2\pm 1.1\text{ at }%
s=17.60\Rightarrow R_0\simeq 198\Rightarrow q_0\simeq -0.0715  \label{19} \\
\text{1}^0 &&\text{: }s_{MIN}=16.5254\text{ at }a_0=22.6\text{; }%
H_0=105.3\pm 0.9\text{ }\Rightarrow R_0\simeq 204\Rightarrow q_0\simeq
-0.0715  \label{20}
\end{eqnarray}
The corresponding unweighed fittings, with $w=1$ given to each group, don't
significantly modify the previous solutions.

\newpage\ 

\subsubsection{Groups by SANDAGE \& TAMMANN}

S\&T sample: 20 nearby Groups of galaxies plus 10 single galaxies from data
summarized by Sandage and Tammann in Table 2 of their paper V (1975). On the
whole these data, when fitted directly with $w$ $=1$ according to the 0$th$
order eq.(6), give a solution which confirms the same Sandage \&\ Tammann $H$
value reported in (8) at page 319 of paper V, that is: 
\begin{equation}
0_{w=1}^0:a_0=1.2\pm 7.3\text{; }H_0=57\pm 5\text{ at }s=18.94\Rightarrow
R_0\approx 0  \label{21}
\end{equation}

Note how the previous unweighed 0$th$ order fitting doesn't produce any
significant dipole, seeing as $a_0=1.2\pm 7.3$, which practically supports
no radius owing to the error oversize.

Otherwise, we can now limit the analysis only to the 20 weighed Groups of
galaxies (187 in all with $\langle r\rangle =14.8$), without the 10 field
single galaxies. Weighing each group by the number of galaxies ($w=n_{obs}$%
), that emerges in Table 1 of the same paper V, the 0$th$ and1$st$ order
weighed fittings of the Group data, here summarized for facility in our
Table 1 where the Virgo Cluster of S\&T-V Table 2 (1975) has weight $w=32$
according to S\&T Table 4 of paper IV (1974), change as follows: 
\begin{eqnarray}
0^0:a_0= &&9.5\pm 3.0;H_0=66.4\pm 1.9\text{ at }s=17.64\Rightarrow R_0\simeq
215\Rightarrow q_0\simeq -0.0477  \label{22} \\
1^0:s_{MIN}= &&17.2997\text{ at }a_0=9.33;H_0=65.1\pm 1.3\Rightarrow
R_0\simeq 220\Rightarrow q_0\simeq -0.0478  \label{23}
\end{eqnarray}

The results (22)(23) are particularly meaningful because they confirm the
model with $w=n_{obs}$ by means of a correct radius $R_0$ evaluation.; on
the other side the corresponding weighed fitting, that has surely more sense
in a group analysis, reverses the result with respect to the unweighed one
of (21), being now $a_0=9.5\pm 3.0$ and $R_0$ about $220$ $Mpc$ from (9).

\subsubsection{ Individual galaxies by SANDAGE\ \&\ TAMMANN}

S\&T sample: 83 individual galaxies from data summarized by Sandage and
Tammann in Table 3 and Table 4 of their paper V (1975) plus the above cited
10 single galaxies of Table 2 (S\&T-V). Here the galaxy sample, whose
average distance results to be about $28.3$ $Mpc$, is distributed along a
wider distance range with respect to that of Groups; so in this case the
least square fitting of the whole of 83 equally weighed galaxies has been
carried out using the 0$th$ , 1$st$ and 5$th$ order eqs. (6)(8)(14), from
which it follows respectively:

\newpage 

\begin{center}
\begin{tabular}{|c|c|c|c|c|c|}
\hline
Order & $s_{MIN}$ & $a_0$ & $H_0$ & $R_0\simeq $ & $q_0\simeq $ \\ \hline
0$^0$ & $27.8638$ & $11.9$ & $72.7\pm 3.1\text{ }$ & $225$ & $-0.0546$ \\ 
\hline
1$^0$ & $25.8267$ & $12.64$ & $69.8\pm 2.8$ & $259$ & $-0.0604$ \\ \hline
5$^0$ & $25.7606$ & $12.66$ & $69.8\pm 2.8$ & $260$ & $-0.0605$ \\ \hline
\end{tabular}
\ \hspace{1.0in}(23)
\end{center}

These three solutions clearly show a fitting precision improvement with the
increase of analysis order.

Finally, considering as a whole the combined sample of 20 weighted Groups of
187 galaxies plus the above 83 individual galaxies, that is to say all the
data of Table 2-3-4 (S\&T-V) together, the resulting solution 
\begin{equation}
\text{1}^0:s_{MIN}=20.3590\text{ at }a_0=11.70;H_0=67.3\pm 1.2\Rightarrow
R_0\simeq 258\Rightarrow q_0\simeq -0.0580  \tag{24}
\end{equation}
still falls into the error range of that obtained by individual galaxies
only, even if the Group analysis may be a bit affected both by the weighing
procedure and by the computing method of the Hubble ratios, which according
to (8) should all be assembled as $\langle \frac{\dot r}r\rangle $.

\subsection{SOLUTIONS BY THE AARONSON et al. CATALOG}

The opportunity to use homogeneous samples of data comes both from the AA1
sample of 308 nearby individual galaxies (Aaronson et al., 1982), and from
another separate sample of 148 galaxies in 10 clusters (AA2), by Aaronson et
al. (1986). This data set, representing a physical whole of
galaxies/groups/clusters, has the great advantage of covering more than $100$
$Mpc$ in depth and of possessing absolute estimates of distances. In fact
the original data listed in Table 3 of the 1982 Aaronson et al. paper, plus
those listed in Table 2 of the successive 1986 work, adopting the
Tully-Fisher relation calibrated by Aaronson et al. (1986, p. 550), permit
the calculation of the galaxy distances and related Hubble ratios, and
consequently of the average $\langle r\rangle $ and $\langle \dot r/r\rangle 
$ of the involved groups and clusters. These Hubble ratios, $\dot r$
including the standard correction of $300\sin l\cos b$, result to be
corrected by the motion of the Sun in the Local Group, in practice due to
galactic rotation, with the standard vector of $300$ $Km/s$ toward $%
l=90^0,b=0^0$(cf. Sandage \& Tammann, 1975); this means we consider Hubble
ratios as seen from our Local Group, or from the Galaxy, the Milky Way, as
being almost motionless within its Group.

So the obtained Hubble ratios refer to the observed Hubble flow.

Now we shall proceed through two steps of numerical analysis, referring to
individual galaxies and normal Groups\TEXTsymbol{\backslash}Clusters,
respectively.

\subsubsection{1$^0$ Step: Individual galaxy solution}

The first check step starts by applying the $0$th order (6), the $1$st order
(8), the $5$th order (14) equations separately, to all the AA1 sample of 308
individual nearby galaxies, whose distance mean is $\langle r\rangle =16.13$ 
$Mpc$, and to the other AA2 sample of 148 more distant individual galaxies,
with $\langle r\rangle \cong 70.69$ $Mpc$ . Both the samples may be
considered homogeneous and rich enough, even if affected by large scattering
in distance. Indeed, being the average individual distance of AA2 about 5
times greater than that of AA1, the involved solutions and their
confrontation take on a special meaning. The least square method applied to
the algebraic system of the 308 and 148 equations (6)(8)(14), respectively,
gives the solutions listed in the small table below, where $s_{MIN}$ is the
minimum standard deviation of the fit corresponding to the listed $a_0$
value, and all the quantities are in Hubble units:

\begin{center}
\begin{tabular}{|l|l|l|l|l|l|l|}
\hline
Sample & Order & $s_{MIN}$ & $a_0$ & $H_0$ & $R_0\simeq $ & $q_0\simeq $ \\ 
\hline
308AA1 & 0$^0$ & $28.6243$ & $16.4$ & $90.6\pm 1.6$ & $200$ & $-0.060$ \\ 
\hline
148AA2 & 0$^0$ & $19.7237$ & $15.4$ & $98.8\pm 1.6$ & $158$ & $-0.052$ \\ 
\hline
308AA1 & 1$^0$ & $27.5725$ & $16.35$ & $88.0\pm 1.6$ & $211$ & $-0.062$ \\ 
\hline
148AA2 & 1$^0$ & $18.0575$ & $16.6$ & $88.7\pm 1.5$ & $211$ & $-0.062$ \\ 
\hline
308AA1 & 5$^0$ & $27.5508$ & $16.3$ & $88.0\pm 1.6$ & $210$ & $-0.062$ \\ 
\hline
148AA2 & 5$^0$ & $18.0072$ & $16.8$ & $88.2\pm 1.5$ & $216$ & $-0.063$ \\ 
\hline
\end{tabular}
\hspace{1.0in}(25)
\end{center}

The above table shows a good solution convergence of $H_0$ and $R_0$ with an
increase of the analysis order, and a corresponding decrease of the standard
deviation absolute value. That is very important and meaningful to the
correctness of the model because of the large difference in distance between
the AA1 and the AA2 samples. Furthermore the table gives clear indication of
the rising trend of $R_0$ still with the order increase.

\subsubsection{2$^0$ Step: Group\TEXTsymbol{\backslash}Cluster solution}

The 2$^0$step solution refers always to the Aaronson et al. catalogue, now
as groups and clusters. Indeed the appropriate Eqs. (6)(8)(14) in two
unknowns, after transforming to represent the average group or cluster
Hubble ratio, should now have $\langle \frac{\dot r}r\rangle ,\langle
r\rangle ,\langle \dot r\rangle ,\langle \cos \gamma \rangle ,\langle r\cos
\gamma \rangle ,\langle r^2\cos \gamma \rangle ,....$ But, having here
verified that the results are unaffected, we have more practically used $%
\cos \gamma _{cluster},$ $\langle r\rangle \langle \cos \gamma \rangle
,\langle r\rangle ^2\langle \cos \gamma \rangle ,...$instead of $\langle
\cos \gamma \rangle _{cluster},\langle r\cos \gamma \rangle ,\langle r^2\cos
\gamma \rangle ,...$

The relative group and cluster data have been reported in Table 2 and Table
3, where 31 nearby groups (139 galaxies) and 11 nearby clusters (164
galaxies) of the Aaronson data set (Aaronson et al., 1982-1986) are listed,
respectively. The quantities tabulated are: the galaxy number for groups and
clusters, $n_{obs}$; the average equatorial coordinates, $\langle \alpha
\rangle $ and $\langle \delta \rangle ;$ the average functions, $\langle
\cos \gamma \rangle _{group}$ or $\cos \gamma _{cluster},$ between the
direction of the Bahcall \& Soneira (1982) void center and that of the group
members or cluster; the group\TEXTsymbol{\backslash}cluster average observed
velocity, $\langle \dot r\rangle ;$ the group\TEXTsymbol{\backslash}cluster
average distance, $\langle r\rangle ;$ and lastly the average Hubble ratio, $%
\langle \frac{\dot r}r\rangle $.

The least square method applied to the algebraic system of the 31 group
(31GR) and 11 cluster (11CL) equations (6)(8)(14), using $n_{obs}$ as
weights in the fitting, gives the solutions listed in the table below, where
in addition 10CL represents the same Aaronson 11 cluster sample without the
nearby Virgo cluster (cf. Table 6 by Aaronson et al., 1986) :

\begin{center}
\begin{tabular}{|l|l|l|l|l|l|l|}
\hline
Sample & Order & $s_{MIN}$ & $a_0$ & $H_0$ & $R_0\simeq $ & $q_0\simeq $ \\ 
\hline
31GR & 0$^0$ & $12.756$ & $18.5$ & $91.34\pm 1.08$ & $222$ & $-0.068$ \\ 
\hline
11CL & 0$^0$ & $4.7733$ & $16.8$ & $98.07\pm 0.37$ & $175$ & $-0.057$ \\ 
\hline
10CL & 0$^0$ & $3.6965$ & $15.4$ & $98.79\pm 0.30$ & $158$ & $-0.052$ \\ 
\hline
31GR & 1$^0$ & $12.395$ & $17.8$ & $88.99\pm 1.05$ & $225$ & $-0.067$ \\ 
\hline
11CL & 1$^0$ & $4.2090$ & $17.1$ & $88.29\pm 0.33$ & $219$ & $-0.065$ \\ 
\hline
10CL & 1$^0$ & $4.3889$ & $16.8$ & $88.42\pm 0.36$ & $215$ & $-0.063$ \\ 
\hline
31GR & 5$^0$ & $12.498$ & $18.16$ & $88.94\pm 1.06$ & $229$ & $-0.068$ \\ 
\hline
11CL & 5$^0$ & $4.3430$ & $17.2$ & $87.96\pm 0.34$ & $222$ & $-0.065$ \\ 
\hline
10CL & 5$^0$ & $4.5501$ & $16.9$ & $88.04\pm 0.37$ & $218$ & $-0.064$ \\ 
\hline
\end{tabular}
\hspace{1.0in}(26)
\end{center}

The most impressive result of table (25) and table (26), is the practical
coincidence between the 308AA1-5$th$ order solution with the 148AA2-5$th$
order one, or with that of the 10CL-5$th$ order sample. The only light
difference refers to the $R_0$ value derived from $a_0$. Table (26) again
confirms the rising trend of $R_0$ with the order increase.

For completeness of the (29) Group analysis, a further 1$st$ order fitting
test on the data listed in Table 2 has been carried out now using as Hubble
ratios those resulting from the division of the average radial velocity $%
\langle \dot r\rangle $ with the average distance $\langle r\rangle $ , that
is $\frac{\langle \dot r\rangle }{\langle r\rangle }$ instead of $\langle 
\frac{\dot r}r\rangle $. Indeed this procedure doesn't seem correct because
of the assumed application of the expansion equation (8) to elementary $%
splinters$ as the individual galaxies are, according to the toy model of
Paper I. But at the same time this method of Hubble ratio calculation is the
only possible one which permits further analyses of far clusters and
superclusters. Consequently here we report, as example, a second 1$st$ order
solution of the 31GR sample, which is: 
\begin{equation}
31AA1GR_{\frac{\langle \dot r\rangle }{\langle r\rangle }}:s=11.35572\text{
at }a_0=17.8\text{; }H_0=86.3\pm 1.0\Rightarrow R_0\simeq 239\Rightarrow
q_0\simeq -0.069  \tag{27}
\end{equation}

Of course all the previous solutions strictly depend on the absolute
calibration carried out by Aaronson et al. in their 1986 paper. It is
interesting to note, only as an example, how a complete coincidence of $H_0$
and $R_0$ of the individual galaxy sample 308AA1 with the equivalent ones by
Sandage \&\ Tammann, that is with the solution (26), may be simply obtained
through an increment of half negative magnitude to the zero-point of the
quadratic IR/H relation (Aaronson et al., 1986-p.550).

\section{GENERAL SOLUTION}

The $q_0$ constancy condition means that the product $H_0R_0$ is invariable
in presence of systematic variation of the distance estimates; so, in order
to discriminate among the above historic samples by de Vaucouleurs, Sandage
\& Tammann and Aaronson et al., we can now directly proceed to test the
above listed different solutions.

Indeed only the individual galaxy samples seem to satisfy sufficiently the
conservation condition of $q_0.$ In fact, it being important to take into
account both the meaningful representation in terms of number of galaxies
and the careful reliability shown by more coinciding solutions of the
Aaronson et al. catalogue, the 308AA1 5$th$ order solution can be rightly
taken as pilot reference solution in our comparison procedure, which, in the
light of the verified deviation of $q_0$, gives acknowledgement only to the
validity of the 83 individual galaxy sample by Sandage \& Tammann (1975).

Consequently to this $q_0$ test, at present, the general solution which
faithfully represents all the previous numerical analysis, may be summarized
by the alternative and separate (23)(25) results, as follows: 
\begin{equation}
Sample-83S\&T\text{ }IG:H_0=70\pm 3;R_0\approx 260\Rightarrow H_0R_0\approx
18200Km/s  \tag{28}
\end{equation}
\begin{equation}
Sample-308AA1\text{ }IG:H_0=88.0\pm 1.6;R_0\approx 210\Rightarrow
H_0R_0\approx 18500Km/s  \tag{29}
\end{equation}

\newpage\ 

\section{CONCLUSIONS}

\subsection{MINI CHECK ATLAS OF HUBBLE RATIO DIPOLES}

(Optional section, now limited to 7 plotted Hubble ratio dipoles (see
Figures 2-3-4-5-6-7-8))

\vspace*{1.0in}

\textbf{REFERENCES}

Aaronson, M. et al. 1982, Ap. J. Suppl. Series, 50, 241

Aaronson, M. et al. 1986, Ap. J. 302, 536

Bahcall, N.A. and Soneira, R.M. 1982, Ap. J. 262, 419

Lorenzi, L. 1996, Astro. Lett. \& Comm., 33, 143

\hspace{1.2in} (1994-Grado3 Proc., SISSA ref. 155/94/A)

\hspace{.7in}1995c, Sesto Pusteria International Workshop Book, eds.-SISSA
ref. 65/95/A

\hspace{.7in}1993, 2nd National Meeting on Cosmology, eds. Mem. S.A.It., V.
66, N. 1-1995

\hspace{.7in}1991, Contributo N. 1, Centro Studi Astronomia-Mondov\`\i -Italy

\hspace{.7in}1999-I, (submitted)

Sandage, A. and Tammann, G.A., 1974, Ap. J. 194, 559

Sandage, A. and Tammann, G.A., 1975, Ap. J. 196, 313

Vaucouleurs, G. de 1965, in ''Galaxies and the Universe'', 1975, Vol. IX of
Stars and

\hspace{1.0in}Stellar Systems, p. 566, The University of Chicago Press

\newpage

\textbf{Table 1}\ 

\vspace{.2in}

\begin{tabular}{|c|c|c|c|c|c|c|c|}
\hline
GROUP & $n_{obs}$ & $L_{SG}$ & $B_{SG}$ & $\cos \gamma $ & $\dot r$ & $r$ & $%
\frac{\dot r}r$ \\ \hline
$G$ $253$ & $9$ & $265$ & $-3$ & $-0.7114$ & $281$ & $3.4$ & $82.6$ \\ \hline
$G$ $672$ & $2$ & $327$ & $-4$ & $-0.0028$ & $540$ & $10.9$ & $49.5$ \\ 
\hline
$G$ $1023$ & $6$ & $342$ & $-9$ & $+0.2579$ & $721$ & $14.3$ & $50.4$ \\ 
\hline
$G$ $1068$ & $6$ & $305$ & $-26$ & $-0.0597$ & $1332$ & $18.1$ & $73.6$ \\ 
\hline
$Eridanus$ $G$ & $22$ & $278$ & $-43$ & $-0.0809$ & $1520$ & $22.8$ & $66.7$
\\ \hline
$G$ $2841$ & $7$ & $50$ & $-16$ & $+0.9336$ & $601$ & $7.6$ & $79.1$ \\ 
\hline
$G$ $2985$ & $2$ & $39$ & $+3$ & $+0.7337$ & $1274$ & $16.1$ & $79.1$ \\ 
\hline
$G$ $M81$ & $10$ & $42$ & $+1$ & $+0.7689$ & $226$ & $3.25$ & $69.5$ \\ 
\hline
$G$ $3184$ & $7$ & $64$ & $-16$ & $+0.9436$ & $673$ & $15.4$ & $43.7$ \\ 
\hline
$G$ $Leo$ & $20$ & $94$ & $-26$ & $+0.8618$ & $799$ & $23.4:$ & $34.1:$ \\ 
\hline
$G$ $3938$ & $5$ & $71$ & $0$ & $+0.8041$ & $873$ & $20.3$ & $43.0$ \\ \hline
$G$ $CVnI$ & $11$ & $69$ & $+6$ & $+0.7447$ & $339$ & $5.0$ & $67.8$ \\ 
\hline
$Virgo$ $CL$ & $32$ & $104$ & $-2$ & $+0.6089$ & $1111$ & $19.8$ & $56.1$ \\ 
\hline
$G$ $Coma$ $I$ & $15$ & $88$ & $+2$ & $+0.7028$ & $922$ & $10.2$ & $90.4$ \\ 
\hline
$G$ $CVn$ $II$ & $7$ & $76$ & $+6$ & $+0.7232$ & $698$ & $7.6$ & $91.8$ \\ 
\hline
$G$ $5128$ & $6$ & $160$ & $-5$ & $-0.0917$ & $317$ & $7.9$ & $40.1$ \\ 
\hline
$G$ $M51$ & $4$ & $72$ & $+17$ & $+0.5985$ & $606$ & $9.7$ & $62.5$ \\ \hline
$G$ $M101$ & $2$ & $64$ & $+23$ & $+0.5281$ & $402$ & $7.2$ & $55.8$ \\ 
\hline
$G$ $6643$ & $8$ & $30$ & $+31$ & $+0.3127$ & $1842$ & $24.4$ & $75.5$ \\ 
\hline
$G$ $IC$ $342$ & $6$ & $11$ & $0$ & $+0.5374$ & $122$ & $4.5:$ & $27.1:$ \\ 
\hline
\end{tabular}

\begin{tabular}{|c|c|c|c|c|c|c|c|}
\hline
GROUP & $n_{obs}$ & $\langle \alpha \rangle $ & $\langle \delta \rangle $ & $%
\langle \cos \gamma \rangle $ & $\langle \dot r\rangle $ & $\langle r\rangle 
$ & $\langle \frac{\dot r}r\rangle $ \\ \hline
$N24/45$ & $2$ & $0^h.158$ & $-24^0.35$ & $-0.74040$ & $553.38$ & $7.92$ & $%
72.64$ \\ \hline
$N134$ & $2$ & $0^h.497$ & $-30^0.8$ & $-0.70834$ & $1581.5$ & $14.47$ & $%
109.68$ \\ \hline
$SCULPTOR$ & $3$ & $0^h.472$ & $-26^0.49$ & $-0.69173$ & $225.55$ & $2.89$ & 
$78.34$ \\ \hline
$N701/755$ & $2$ & $1^h.854$ & $-9^0.63$ & $-0.33592$ & $1766.9$ & $15.83$ & 
$111.72$ \\ \hline
$N1023$ & $5$ & $2^h.395$ & $+32^0.4$ & $+0.14833$ & $1053.1$ & $9.98$ & $%
102.59$ \\ \hline
$ERIDANUS$ & $8$ & $3^h.288$ & $-24^0.21$ & $-0.14530$ & $1478.7$ & $14.66$
& $104.40$ \\ \hline
$FORNAX$ & $7$ & $3^h.523$ & $-34^0.16$ & $-0.18237$ & $1406.0$ & $13.58$ & $%
107.70$ \\ \hline
$N2403-M81$ & $6$ & $8^h.648$ & $+69^0.09$ & $+0.71693$ & $186.28$ & $2.52$
& $82.28$ \\ \hline
$N2336$ & $3$ & $7^h.228$ & $+81^0.57$ & $+0.60823$ & $2349.7$ & $25.38$ & $%
93.00$ \\ \hline
$N2841$ & $3$ & $8^h.697$ & $+51^0.35$ & $+0.92511$ & $681.90$ & $11.91$ & $%
58.11$ \\ \hline
$N3079/U5459$ & $2$ & $10^h.029$ & $+54^0.62$ & $+0.89064$ & $1193.6$ & $%
16.16$ & $73.94$ \\ \hline
$N3184$ & $3$ & $10^h.571$ & $+41^0.54$ & $+0.92196$ & $680.92$ & $12.10$ & $%
56.70$ \\ \hline
$LEO$ & $5$ & $11^h.064$ & $+14^0.26$ & $+0.83873$ & $1016.3$ & $16.54$ & $%
62.54$ \\ \hline
$N3521$ & $3$ & $10^h.92$ & $+2^0.07$ & $+0.77571$ & $808.03$ & $14.57$ & $%
59.09$ \\ \hline
$LEO$ $TRIPLET$ & $3$ & $11^h.287$ & $+13^0.5$ & $+0.81236$ & $682.83$ & $%
8.50$ & $81.26$ \\ \hline
$URSA$ $MAJOR$ & $24$ & $11^h.857$ & $+49^0.29$ & $+0.79018$ & $1011.88$ & $%
14.92$ & $69.20$ \\ \hline
$COMA$ $I$ & $6$ & $12^h.451$ & $+29^0.29$ & $+0.70951$ & $994.54$ & $11.15$
& $98.16$ \\ \hline
$VIRGO$ $SOUTH$ & $7$ & $12^h.635$ & $+1^0.4$ & $+0.51218$ & $1071.9$ & $%
13.50$ & $84.04$ \\ \hline
$CVn$ $I$ & $3$ & $12^h.477$ & $+35^0.87$ & $+0.70501$ & $397.41$ & $4.38$ & 
$90.90$ \\ \hline
$N5033$ & $3$ & $13^h.211$ & $+36^0.7$ & $+0.61212$ & $931.57$ & $13.14$ & $%
70.94$ \\ \hline
$M51$ & $2$ & $13^h.196$ & $+43^0.3$ & $+0.62942$ & $532.12$ & $6.93$ & $%
76.82$ \\ \hline
$M101$ & $3$ & $13^h.883$ & $+56^0.59$ & $+0.55330$ & $364.51$ & $3.66$ & $%
111.17$ \\ \hline
$N5371$ & $3$ & $13^h.747$ & $+41^0.78$ & $+0.54119$ & $2664.4$ & $26.69$ & $%
101.23$ \\ \hline
$N5364$ & $2$ & $13^h.812$ & $+4^0.84$ & $+0.30626$ & $1267.3$ & $14.02$ & $%
102.54$ \\ \hline
$N5566$ & $9$ & $14^h.626$ & $+1^0.68$ & $+0.09895$ & $1617.9$ & $20.03$ & $%
84.06$ \\ \hline
$N5676$ & $3$ & $14^h.508$ & $+49^0.99$ & $+0.45445$ & $2342.8$ & $29.94$ & $%
79.33$ \\ \hline
$N5866$ & $2$ & $15^h.192$ & $+56^0.85$ & $+0.39474$ & $908.10$ & $11.32$ & $%
81.58$ \\ \hline
$N6070$ & $3$ & $15^h.993$ & $+0^0.54$ & $-0.21742$ & $2032.4$ & $26.80$ & $%
75.83$ \\ \hline
$GRUS$ & $8$ & $22^h.895$ & $-40^0.47$ & $-0.90251$ & $1614.1$ & $16.20$ & $%
104.07$ \\ \hline
$N7320/7331$ & $2$ & $22^h.571$ & $+33^0.92$ & $-0.37960$ & $1079.3$ & $%
10.80 $ & $104.02$ \\ \hline
$N7537/7541$ & $2$ & $23^h.202$ & $+4^0.24$ & $-0.68711$ & $2865.7$ & $27.08$
& $106.60$ \\ \hline
\end{tabular}

\newpage\ 

\textbf{TABLE 3}

\vspace{.2in}

\begin{tabular}{|c|c|c|c|c|c|c|c|}
\hline
\hspace{.1in}CLUSTER\hspace*{.1in} & $n_{obs}$ & $\alpha $ & $\delta $ & $%
\cos \gamma $ & $\langle \dot r\rangle $ & $\langle r\rangle $ & $\langle 
\frac{\dot r}r\rangle $ \\ \hline
$PISCES$ & $20$ & $1^h$ & $+30^0$ & $-0.1250$ & $5274$ & $52.91$ & $102.77$
\\ \hline
$A400$ & $7$ & $2^h.917$ & $+5^0.83$ & $+0.0320$ & $7855$ & $82.08$ & $98.02$
\\ \hline
$A539$ & $9$ & $5^h.233$ & $+6^0.38$ & $+0.5306$ & $8536$ & $95.94$ & $91.33$
\\ \hline
$CANCER$ & $22$ & $8^h.3$ & $+21^0.23$ & $+0.9748$ & $4789$ & $61.15$ & $%
83.07$ \\ \hline
$A1367$ & $20$ & $11^h.7$ & $+20^0.12$ & $+0.7903$ & $6486$ & $76.04$ & $%
89.74$ \\ \hline
$COMA$ & $13$ & $12^h.95$ & $+28^0.25$ & $+0.6267$ & $7310$ & $80.72$ & $%
92.46$ \\ \hline
$Z74-23$ & $13$ & $14^h$ & $+9^0.57$ & $+0.3041$ & $5939$ & $72.86$ & $84.34$
\\ \hline
$HERCULES$ & $11$ & $16^h.05$ & $+17^0.93$ & $-0.0697$ & $10733$ & $114.05$
& $96.65$ \\ \hline
$PEGASUS$ & $22$ & $23^h.3$ & $+7^0.92$ & $-0.6380$ & $4275$ & $40.89$ & $%
111.15$ \\ \hline
$A2634/66$ & $11$ & $23^h.67$ & $+24^0$ & $-0.4027$ & $8693$ & $86.23$ & $%
102.38$ \\ \hline
$VIRGO$ & $16$ & $12^h.476$ & $+12^0.34$ & $+0.6243$ & $1064$ & $14.91$ & $%
78.00$ \\ \hline
\end{tabular}

\end{document}